# Geo-Conquesting Based on Graph Analysis for Crowdsourced Metatrails from Mobile Sensing

Bo-Wei Chen and Wen Ji

*Abstract*—This article investigates graph analysis for intelligent marketing in smart cities, where metatrails are crowdsourced by mobile sensing for marketing strategies. Unlike most works that focused on client sides, this study is intended for market planning, from the perspective of enterprises.

Several novel crowdsourced features based on metatrails, including hotspot networks, crowd transitions, affinity subnetworks, and sequential visiting patterns, are discussed in the article. These smart footprints can reflect crowd preferences and the topology of a site of interest. Marketers can utilize such information for commercial resource planning and deployment. Simulations were conducted to demonstrate the performance. At the end, this study also discusses different scenarios for practical geo-conquesting applications.

## I. INTRODUCTION

AS the advancement in communication and computing technologies stimulates the progress of smart cities, trillions of sensors are deployed in every corner of a city, subsequently forming a huge sensor network. Data collected by various sensors range from buildings, streets, and transportation systems, to natural spaces. These heterogeneous data respectively delineate the characteristics of a city. Among these sensors, mobile sensing is becoming popular due to its mobility and pervasiveness. Furthermore, it reflects human social behavior and shows the interaction between persons and a city.

According to a survey conducted by www.statisticbrain.com, the total number of worldwide cellular phone subscriptions was as high as 6.9 billion in 2014. The total app downloads for *i*OS and Android smartphones respectively reached 29 billion and 31 billion.

With such a tremendous number of subscribers using a variety of apps, crowdsourced data from mobile sensing become a valuable resource for market planning. For marketers, user behavior reasoning is an important subject for targeting potential customers. Analysis on mobile sensing data provides different clues about user dynamics. One of the feasible analyses is geo-conquesting.

Geo-conquesting is a newly emerging technology in computational advertising. Such a phrase originates from a combination of two words — Geography and conquesting. According to the definition of conquesting mentioned in [1], it means to deploy an advertisement next to competitors or the products of rivals. Recent location-aware technologies, such as geo-locationing and geo-fencing, have upgraded the scale of conquesting to cyber-enabled geo-conquesting. With outdoor and indoor positioning, for example, *i*Beacon and IndoorAtlas, user locations in geographical areas of interest can be pinpointed more accurately than before. Marketers can use geo-conquesting to hyper-target consumers proactively. Ambient services or products are fed back in real time to users, depending on their locations. This success is attributed to advances in portable and sensing technologies, which have brought traditional advertising industry into a new era. Conventional demographics (e.g., residential density and population density) no longer catch up with city dynamics. Demographics are static, but geo-conquesting counts on live dynamic data.

Geo-conquesting is not merely smart advertising. It represents proactive market strategic planning. No advertisement spamming is involved. Take retail store locationing for example. Selecting a store site is the decisive factor in profit making. Storeowners know that large traffic flows along the side of a road dominate in-store visiting. However, which side should they choose? What type of stores should they operate? The observation in Fig. 1 reveals an interesting example. We use convenience stores, 7-11, as a case study. The area in Fig. 1 is 1.62 km × 1.20 km, and there are 11 stores. The finding shows that eight stores are on the right-hand side. The explanation is that when people leave their home and head for the downtown, they have higher chances to visit convenience stores if they temporarily need something. As opposed to convenience stores, business operators of supermarkets can select the site based on such a principle. When people head for suburban areas back to their houses, hypermarkets that offer daily necessities attract more customers.

B.-W. Chen is with the School of Information Technology, Monash University, Australia

W. Ji is with the Institute of Computing Technology, Chinese Academy of Sciences, China

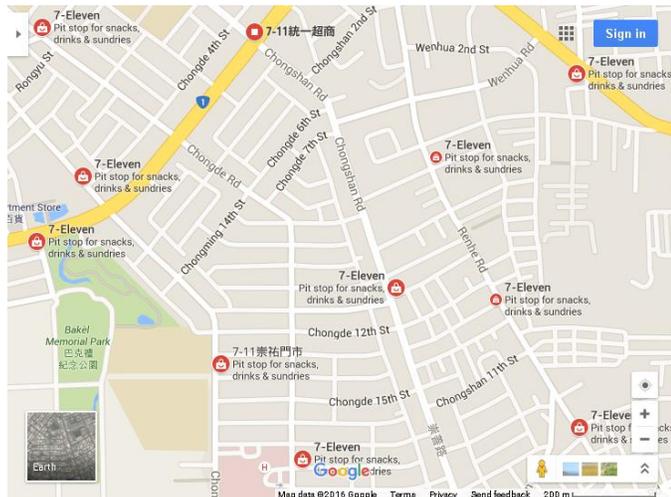

Fig. 1. Example of geo-conquesting by monitoring in-town and out-of-town flows. The direction of downtown is northwest. Eight out of 11 convenience stores, marked as red circles, are located on the right-hand side. Notably, cars drive by the right side of the road in this city.

In the cyber age, metadata collected via mobile sensors facilitate geo-conquesting, for mobile sensing reveals customer activities and environmental conditions. Marketers can benefit from these crowdsensed data by carefully investigating regional dynamics.

Crowd behavior can be further manifested when mobile crowdsensing is applied. Based on communication types [2], mobile sensing can be classified into two categories — Direct and indirect sensing. The former involves direct communication between devices (e.g., device-to-device and machine-to-machine approaches) or direct communication between terminals and base stations (e.g., radio networks). Signals during transmission can be directly used for power management and resource planning, subsequently utilized for estimating the number of cellular phone subscribers or user densities. The latter, indirect sensing, relies on intermedia, e.g., social websites and clouds, where the captured data via mobile sensors can be downloaded by analysts and further processed.

For market analytics, indirect sensing is more convenient than direct sensing because the indirect mechanism avoids negotiations with wireless service providers and telecommunication companies. Data can be accessed via dedicated apps or harvested from social websites.

To analyze crowd behavior, urban monitoring is a feasible approach for sampling the dynamics of a city. Relevant research on mobile crowdsensing, like [3, 4], used vehicular sensors to collect the characteristics of a city. Lee *et al.* [3] proposed a vehicular sensor network "MobEyes" for gathering information in an urban environment. Each vehicle represented a sensor node and carried standard equipment for sensing, communication, and storage. Their system was capable of generating street images, recognizing license plates, and sharing information with other drivers and the police. Zheng *et al.* [4] analyzed the traces, generated by the Global Positioning Systems (GPSs) in taxicabs, to detect flawed urban planning in a city, such as traffic problems in the regions of interest. They used taxi trajectories and the partitioned regions of a city for modeling passenger transitions. Frequent transition patterns were extracted by mining a large transition graph. Transitional time was examined for finding flawed regions. The advantage of vehicular sensing is that the hardware specifications of sensors are usually better than those of mobile phones. This provides better quality during data collection. However, the routes of vehicles are fixed, and the traces are limited to the places accessible to vehicles.

Reades *et al.* [5] developed a measure of bandwidth usage for radio networks, called Erlang. A single unit of Erlang was defined as one person-hour of phone use. They monitored the Erlang data in a region of 47 $km^2$ over four months as these data could provide bandwidth consumptions and insights into the spatial and temporal dynamics of a city. By observing accumulated Erlang data, salient regions were revealed on the map. Like [5], Calabrese *et al.* [6] also employed cellular phones as the media for urban monitoring. The researchers developed the Localizing and Handling Network Event System (LocHNES) to monitor traffic conditions and pedestrian/vehicle movements. The LocHNES employed the radio network database, the antenna database, and the signal-propagation model, all provided by a local telecommunication company, to locate subscribers. When events occurred, e.g., call-in process, message sending, and handover, the system was capable of tracking subscribers. As mentioned earlier, urban monitoring based on radio networks requires participation of telecommunication companies. For third-party marketers, e.g., advertising agency, such data need purchasing, and this increases costs.

Based on a recent survey carried out by www.statisticbrain.com, more and more smartphone users prefer using apps for communication, so urban sensing via mobile apps becomes another way for collecting data. Furthermore, the captured data can be forwarded and uploaded to cloud sides stealthily without interfering with user operations. Closely examining the metadata crowdsourced from mobile users shows different demands of customers. This is beneficial for both sides — Marketers and customers.

Related innovations were frequently published in literature. For example, Lu *et al.* [7] devised a pattern recognition approach for acoustic signal processing. The signal, including human voices and ambient sound, was collected by using phone receivers. Typical classifiers were used to recognize sound events. Likewise, Xu *et al.* [8] also used machine learning to analyze sounds, but their application focused on speaker counting. The systems respectively developed by Kanjo [9] and by Rana *et al.* [10] were designed for monitoring the noise level of a city by using mobile phones. Actually, these aforementioned creative systems can be furthered to monitor the background sound, e.g., babble, by analyzing phone conversations. With GPS trails, marketers are capable of learning the location information of the crowds. Direct marketing becomes effective.

Rather than using acoustic information, [11] and [12] concentrated on images collected from mobile social networks. The former discovered the people, activity, and context (e.g., indoor/outdoor and location names) in a picture, whereas the latter involved semantic interpretation and understanding on flyers. Both adopted machine learning and social network analyses. Compared with acoustic signals, images tell more tales than audio does. For instance, crowdsourced photographs taken in a place and at a certain moment probably indicate a tourist attraction or a popular social activity. With geospatial and temporal analyses, marketers can publicize their products easily. Such methods as sales-force allocation technology for electronic advertisement insertion or retail location analysis for physical store deployment can respectively enhance the product image.

At present, much effort has been devoted to the research on consumer purchasing behavior and patterns. Nevertheless, few IT approaches were proposed for maximizing the interest of enterprises. To this end, this study concentrates on the side of marketers. In the rest parts, several novel ideas are examined to deal with the challenge of geo-conquesting in smart cities.

## II. Crowdsourced Features

Customer trails, as mentioned earlier, provide useful clues for market analysis. Previous works [5, 6, 13] usually focused on generation of heatmaps for displaying regions of interest in a city. However, valuable information is hidden in the crowdsourced trails, for instance, sequential visiting patterns, transition flows, and site combinations. To extract such hidden information that reflects customer behavior, this study investigates the dynamics behind crowdsourced trails, including hotspot networks, transitions, and affinity subnetworks by using graph analysis. Figure 2 shows the overview. Heterogeneous data transmission while different apps are running is used in the model. Such information is another indicator of user dynamics, especially when the data are co-displayed with trajectories. Typically, two types of data are formed while apps are running. One is the trails based on app types (e.g., instant messaging apps), and the other is crowdsensed multimedia. This study highlights the former type rather than the latter because crowdsensed multimedia involve pattern recognition, and this study focuses on graph analysis.

As these data are generated along with trails, app types and their timestamps are labeled in the trajectories. We use the term "metatrails" to represent them. Metatrails contain more geospatial and temporal characteristics that feature the preference of customers than GPS trails do. Different metatrails render various activities of customers. More importantly, metatrails are embedded with mobile geosocial networks.

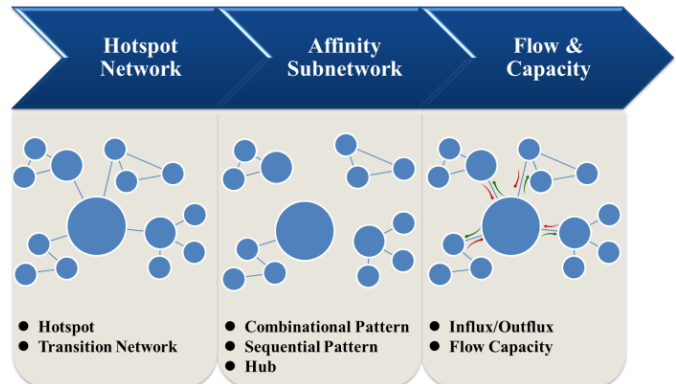

Fig. 2. Illustration of the system

### A. Hotspot Network

This feature is inspired by GPS navigation systems, widely used in our daily lives. For a tourist attraction, it is usually a frequently visited spot, or a finish point. Observations show that a waypoint or the start point on the route is sometimes another frequently visited spot. With sufficient data, there exists a frequently visited route between two hotspots. Large-scale metatrails collected from crowds can substantiate such an observation. This feature is applicable in a city or a site. The following steps show how to build a hotspot network.

Let an edge denote a road, and let a vertex represent a point of interest (i.e., a waypoint, a start point, or a finish point). According to graph theory, a walk is defined as a sequence of alternating vertices and edges. Let us also define the distance between two directly connected vertices as one.

Given a frequently visited spot $r$ on a map, a tree is created by tracing all the routes (i.e., walks), of which the distance is one. Next, removing all the vertices, of which the visiting frequencies are lower than a predefined threshold, generates a new pruning tree. Iteratively selecting a vertex in this tree yields a network $G$.

Based on all the vertices and edges in this network, a matrix of transition probabilities is formed by calculating the in-degrees and the out-degrees of the vertices. Notably, a user trajectory is a sequence of coordinates with timestamps. Therefore, the system can compute the in-degree and the out-degree of a vertex. For example, assuming there are three vertices $v_2$–$v_4$ adjacent to vertex $v_1$, the out-degrees from $v_1$ to $v_2$–$v_4$ are respectively one, two, and three. Thus, transition probabilities from $v_1$ to $v_2$–$v_4$ are respectively 1/6, 2/6, and 3/6.

Compared with heatmaps that display frequently visited areas without showing connections between them, hotspot networks use transition probabilities to present user preferences and flows between places.

An example of hotspot networks is shown in Fig. 3. This hotspot network displays frequently visited stores (i.e., red marks) in a mall, where the black lines are frequently visited routes.

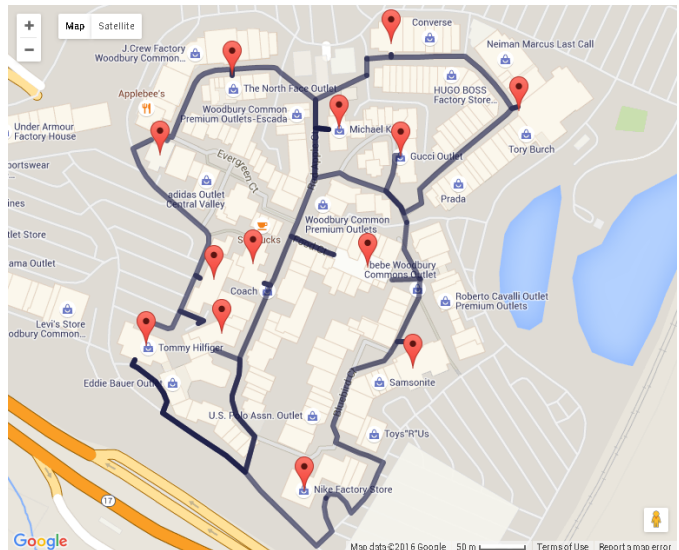
Fig. 3. Example of hotspot networks (generated by gmaps.js): Mall

### B. Affinity Subnetworks — Combo-Site Mining

The transition matrix of a hotspot network represents the preference of mobile crowds when they move between places. A visiting pattern can be discovered by performing subgraph analysis. One feasible way is graph clustering. When graph clustering is applied to a hotspot network, hotspots are grouped together, subsequently forming several subnetworks. As the objective of graph clustering is to group vertices that present high connectivity, we denote the resulting subnetworks as "affinity subnetworks." The affinity herein is used to describe crowd preferences. Affinity subnetworks further the analysis for retail site selection from single sites to combo-site selection. Combo sites take multiple effects and yield more profits than single sites.

At present, many graph clustering approaches have been proposed. Among these approaches, spectral clustering, Markov clustering, minimum cut, and $K$-means are most related to our case. Notably, crowdsourced data can reach trillions. To process billions of vertices in a network, complexity is of prior concerns. Algorithms that involve matrix decomposition like Principal Component Analysis and Singular Value Decomposition create too much computational time. Thus, they are inappropriate in our case. In the following content, we use Markov clustering as a case study because it is directly related to transition probabilities.

Markov clustering was derived from flow simulation by Stijn van Dongen [14]. It used the ideas of Markov chains and random walks within a graph by iteratively computing transition matrices based on edge weights. The intuition behind Markov clustering is that if a graph possesses a clustered structure, random walks between vertices lying in the same cluster are more likely than those between vertices which are located in different clusters [14]. This finding is based on the equilibrium distribution of Markov chains. Let $\pi$ represent the matrix of initial probabilities for all the vertices in a hotspot network $G$. By multiplying $\pi$ by a transition-probability matrix $\mathbf{T}$ within finite times, the resulting product becomes stable. Furthermore, regardless of start points, the equilibrium distribution is the same. Markov clustering employs two major operations — Expansion and inflation — for graph clustering. The former tests connectivity between vertices when taking the Hadamard product. The latter increases tightness of clusters. Eventually, iterations result in separation of the network.

After clustering, each cluster forms an affinity subnetwork, namely, a combination of frequently visited hotspots, as shown in Fig. 4. An affinity subnetwork indicates that people prefer visiting these places in combination. This is because Markov clustering simulates people randomly walking in these hotspots based on their interest (i.e., transition probabilities) when sufficient crowdsourced trails are collected. Therefore, strong connectivity is created among hotspots.

A hub in an affinity subnetwork, i.e., a vertex with the highest degree, is a pivotal place which people at adjacent hotspots have higher chances to visit. Mining affinity subnetworks and hubs helps marketers discover combo stores and combo hotspots.

Each affinity subnetwork can be contracted to a vertex. Thus, connections between clusters are easily visualized.

### C. Sequential Visiting Pattern — Between Hotspots and Between Subnetworks

Mining the sequential patterns of mobile footprints is conducive to predicting crowd preferences and arranging store locations. Nevertheless, it is difficult to analyze large-scale trails because of complexity. Besides, sequential patterns with long duration are not practical for market planning. Fortunately, affinity subnetworks reduce computational time since graph clustering breaks a large graph into small components.

Unlike affinity subnetworks that concentrate on combo stores, this feature highlights the order of patterns. There are two types of sequential visiting patterns. One is the order of the hotspots in an affinity subnetwork, and the other is the order of subnetworks.

To analyze the order of hotspots, firstly all the trails in an affinity subnetwork are extracted. Then the system rearranges the vertices based on their latest timestamps in the trails. When more than one sequential pattern is generated, it means at least two types of orders exist in an affinity subnetwork.

When the traversal order of different subnetworks is focused, an entire subnetwork is viewed as a vertex. Namely, all the vertices are contracted to generate one vertex. The whole hotspot network $G$ becomes a new graph, of which each vertex represents an affinity subnetwork. The approach for analyzing the order of subnetworks is similar to that for the order of hotspots.

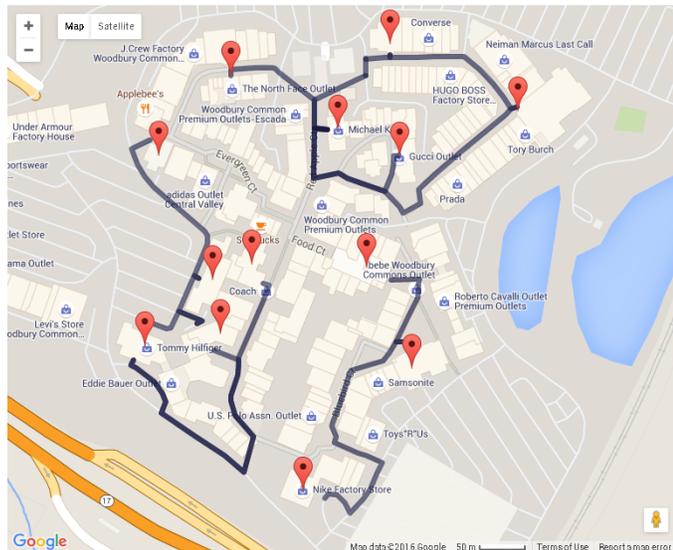

Fig. 4. Example of affinity subnetworks. Three subnetworks were discovered by graph mining when crowd dynamics were considered. Each represents a combination of frequently visited hotspots. People prefer staying in the same subnetwork and visiting these places in combination.

### D. Flows and Capacities of the Hotspot Network

In-store visits involve two factors. One is the directions of traffic flows, and the other is the flow capacity. The former dominates store types that are highly related to customer preferences when they move, as mentioned earlier in Fig. 1. The latter increases store visibility if the location of a store is right next to a major flow. Therefore, monitoring flow directions and flow capacities in a hotspot network is important. Flow directions can be easily manifested by computing in-degrees and out-degrees. However, flow capacities require a more complicated mechanism because both overflows and underflows in a route have influences on in-store visits and visibility. This relatively affects site rentals, product prices, and profits.

The information on the maximum and the minimum flow capacity is useful when marketers select sites. Their requirements for maximum and minimum capacities can be converted into upper and lower bounds, i.e., constraints, during profit optimization. Assume that low flows bring few in-store visits and subsequently low rentals. There is a balancing relation between flows and site rentals. However, this is still a challenge of multiside profit optimization.

### III. Geo-Conquesting

Unlike many works that focused on the user side by mining their transaction records, browsing histories, and social connections, geo-conquesting concentrates on the marketer side. The following section discusses two different scales of geo-conquesting strategies by using the crowdsourced features in the previous section. One is from the perspective of mall management, and the other is for metropolitan planning.

For mall operators, as hotspot networks display frequently visited stores, the difference in rentals between stores with high and low in-store visits can be dynamically adjusted. When rentals decrease, the cost is reflected on the price of retail products, subsequently attracting more customers. There is a mutual interest among mall operators, storeowners, and customers.

Traffic flow management is another benefit brought by hotspot networks. Mall operators can examine transition probabilities, crowd directions, and flow capacities to redeploy mall facilities and balance flows. Compared with crowd flows or heatmaps in the unstructured open space, flows detection based on hotspot networks generate more concrete information on store connectivity. As various app-users generate different routes, electronic billboards or digital billboards can be set up in the hallway for displaying advertisements. Additionally, advertising content can be proactively changed based on flow types for targeting customers. Mall operators can dynamically charge advertising agencies according to flow amounts. When combined with sequential visiting patterns, deployment of stores and digital billboards can be reshaped to fit the flow directions, subsequently bringing in more customers.

For mall operation, how to select a group of stores and deploy them together is an important topic since an appropriate combination of stores creates a weighting factor in in-store visits. With the use of graph clustering and transition flows, the entire hotspot network is separated into several affinity subnetworks. Each subnetwork represents combo stores. With such combo information, mall managers can preallocate space for lease. Storeowners can join an affinity subnetwork and open a store that fits this subnetwork.

For conquesting in a city, the main focus is on profitable regions of interest or points of interest. As large traffic flows increase visibility of stores, a hotspot network provides a good indicator for retail store locationing. Moreover, affinity subnetworks can further store locationing analysis from individual sites to combo-site selection. Advertising agencies can utilize affinity subnetworks to project related-product images onto customers because affinity subnetworks represent customer preferences. When geo-conquesting meets city planning, it becomes city marketing. City planners can improve public transportation and infrastructures by exploring affinity subnetworks. Metropolitan branding will become more effective via dynamic crowdsourcing.

### IV. Numeric Result

As the combo-site analysis is important to geo-conquesting, we conducted a simulation to test the performance of generating affinity subnetworks. A directed weighted graph was randomly synthesized with 500 vertices. The sparse ratios were from 10% to 90% with a separation of 20%. Two typical algorithms for rapid graph clustering — Markov clustering and $K$-means — were benchmarked. Both involved no matrix decomposition. The former used the source codes (developed by Daniel A. Spielmanat, Yale University) with our modifications. The latter followed the algorithm in [15]. Both the iterations were fixed at 500. The Hadamard power was two, and the number of clusters in $K$-means was 10.

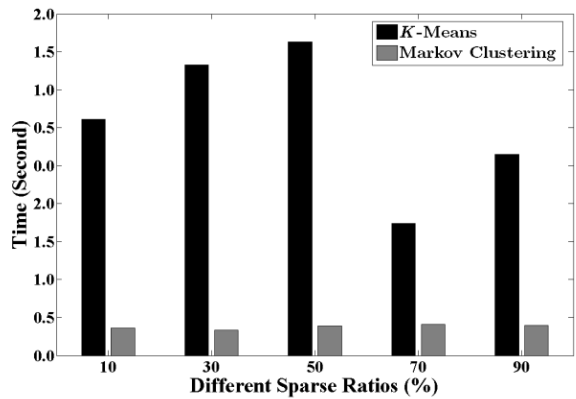

Fig. 5. Graph clustering for the combo-site analysis

Figure 5 shows the computational performance. The vertical axis signifies the computational time, whereas the horizontal axis displays the two different approaches applied to various data. As shown in Fig. 5, when Markov clustering was used, the performance was higher that of $K$-means. The computational time of Markov clustering was 0.1518 second on average, faster than that of $K$-means, 1.2365 seconds. This is conducive to the data analysis as the computational time is favorable when crowdsourced data are collected.

## V. CONCLUSION

This study examines crowdsourced features generated from mobile metatrails for geo-conquesting. To reflect the dynamics of a city, graph analysis is used for discovering hotspot networks, transition probabilities, and flow capacities. Subsequently, affinity subnetworks and sequential visiting patterns are extracted by using rapid graph clustering. Affinity subnetworks (i.e., combo stores) and sequential patterns allow marketers to analyze crowd sequential activities — Between stores and even between subnetworks. Such a discovery creates a weighting factor to in-store visits.

Different scales of geo-conquesting strategies, ranging from mall operations to metropolitan business targeting, are also presented in the discussion. With these features, intelligent marketing becomes feasible because store locationing is based on dynamic city characteristics instead of static demographics.

Interesting challenges arise in intelligent marketing, for example, combo-site locationing, sales-force allocation theory, and multiside profit optimization. In the future, there will be systematic Mathematical equations for modeling these challenges.